\documentclass[prc,twocolumn,showpacs,amsmath,amssymb,superscriptaddress,floatfix,nofootinbib]{revtex4}
\usepackage{mathrsfs,bm}
\usepackage{longtable,lscape}
\usepackage{txfonts}
\usepackage{amssymb}
\usepackage{indentfirst}
\usepackage{graphicx,booktabs}
\usepackage{multirow}
\usepackage{color}
\usepackage{amssymb}
\usepackage{subfigure}

\begin{document}
\title{The production of neutral $N^*(11052)$ resonance with hidden beauty from $\pi^-p$ scattering}

\author{Chen Cheng}
\affiliation{Institute of Modern Physics, Chinese Academy of Sciences, Lanzhou 730000, China}

\author{Xiao-Yun Wang}
\thanks{Corresponding author: xywang@lut.cn}
\affiliation{Department of physics, Lanzhou University of Technology, Lanzhou 730050, China}

\begin{abstract}

We investigate the discovery potential of the predicted neutral hidden beauty $N^*(11052)$ resonance through $\pi^- p$ scattering within an effective Lagrangian approach. Two reactions $\pi^-p\rightarrow K^-\Sigma^+$ and $\pi^-p\rightarrow \eta_bn$ are studied in this work, with nucleon pole exchange as the background. It is found that the contributions of the $N^*(11052)$ resonance give clear peak structures in the magnitude of 1 $\mu b$ near the threshold of the $N^*(11052)$ in the total cross sections. The numerical results indicate that the center of mass energy $W\simeq$ 11-11.1 GeV would be a best energy window for searching the $N^*(11052)$ resonance, where the $N^*(11052)$ signal can be easily distinguished from the background. The COMPASS experiment at CERN's Super Proton Synchrotron (SPS) with pion beam of $\simeq$ 280 GeV will be an ideal platform for searching the super-heavy resonance with hidden beauty, which is hopeful to test the theoretical results.
\end{abstract}

\pacs{14.20.Mr,13.75.Gx,13.30.Eg}
\maketitle

\section{INTRODUCTION}

Searching and explaining the exotic states, which beyond the scheme of conventional quark model of 3-quark $(qqq)$ and quark-antiquark $(q\bar q)$, have became a very intriguing issue in hadron physics. It provides us a good chance to better understand the strong interaction governed by Quantum chromodynamics (QCD). While the conventional quark model~\cite{prd18,prd34}, first introduced by Gell-Mann and Zweig, gives a good description to various ground hadrons, it fails to be consistent with the observations of various excited hadron states.

For example, many charmonium-like and bottomonium-like states XYZ have been observed~\cite{csb59,epjc71} in recent years. Such states definitely cannot be accommodated by the frame of conventional $c\bar c$ or $b\bar b$ states, they are considered as promising multiquark states or molecular states. On the baryon side, the mass order reverse problem between $N^*(1535)$ and $N^*(1440)$ has been long standing in the conventional quark model~\cite{ppnp45}, Since the $N^*(1535)$ resonance with parity $(\frac{1}{2})^-$ being considered the lowest orbitally excited nucleon is expected to be lighter than the first radically excited nucleon $N^*(1440)$. Moreover in the BES experiment the $N^*(1535)$ resonance is observed to have a large coupling to $K\Lambda$ and $\eta N$ channels with strangeness~\cite{npa675,plb510,pn16,ijmpa20} in the $J/\psi\rightarrow\bar pp\eta$ and $J/\psi\rightarrow\bar pK^+\Lambda+c.c.$ reaction. In order to solve the problems above, pentaquark models with hidden strangeness ($qqqs\bar s$) are introduced to describe the exotic state~\cite{prl96,prc77,npa669,npa835}. However, because of the mixture of pentaquark and 3-quark exists in baryons, and there are always some adjustable ingredients in each model, it will be difficult to pin down the nature of $N^*(1535)$. One way to eliminate the ambiguity is to extend the study of hidden strangeness baryons to hidden charm and hidden beauty baryons, they can be easily distinguished from the conventional quark model for their super large mass.

Actually, many works have been done in the hidden charm region. In Refs.~\cite{prl105,prc84}, the $U_{\chi}PT$ approach is used, several $N^*$ and $\Lambda^*$ resonances with hidden charm were dynamically generated in the $PB$ and $VB$ channels, all of them have masses above 4 GeV and widths smaller than 100 MeV. Here $PB$ and $VB$ stand for pseudoscalar meson-baryon and vector meson-baryon, respectively. Several possible reactions were proposed to look for these predicted resonances, including the $\gamma p\rightarrow J/\psi p$ reaction~\cite{prl105,prc84,jpg41}, and meson-proton collision~\cite{epja51,epl109}. Fortunate and delighted enough, two resonances consistent with pentaquark states were observed in $\Lambda^0_b\rightarrow J/\psi K^-p$ decays by the LHCb Collaboration~\cite{prl115} later in 2015. The two newly discovered resonances were named $P_c(4380)^+$ and $P_c(4450)^+$, which have masses of 4380 MeV and 4450 MeV and widths of 205 MeV and 39 MeV. They are probably the partners of the predicted $N^*_{c\bar c}$ resonances in Refs.~\cite{prl105,prc84}. Different methods were proposed to detect the newly observed $P_c$ states for further confirmation, including photoproduction~\cite{prd92} and $\pi^-p$ scattering~\cite{prd93}. Since many works in the hidden charm region have been done, it motivates us to take a step further, searching for the hidden beauty resonances.

Actually, two charged Bottomoniumlike resonances had been reported by the Belle Collaboration ~\cite{prl108} in $\Upsilon(5S)$ decays, with masses of 10607 $\pm$ 2.0 MeV and 10652.2 $\pm$ 1.5 MeV, and widths of 18.4 $\pm$ 2.4 MeV and 11.5 $\pm$ 2.2 MeV. Charged $\pi$ meson in the final states suggests that the minimal quark content of the two newly reported Bottomoniumlike resonances is a four quark combination, and their low widths are consistent with the properties of the super-heavy nucleon resonances predicted in Ref.~\cite{plb709}. In Ref.~\cite{plb709}, the meson-baryon coupled channel unitary approach with local hidden gauge formalism is extended to the hidden beauty region. Two $N^*$ and four $\Lambda^*$ resonances were predicted, with masses all around 11 GeV and widths only a few MeV. Relevant informations of the two $N^*$ are listed in Table~\ref{decay-width}. Their super large masses and very narrow widths result from the $b\bar b$ components in the states, all decays involve the exchange of a heavy beauty vector, hence are suppressed. According to the informations listed in Tab.~\ref{decay-width}, we found it would be an ideal channel to search and study the neutral $N^*(11052)$ resonances through the $\pi^- p$ scattering, with productions of $K^-\Sigma^+$ or $\eta_b n$ because of their dominant decay branch ratios. And since $K^-$ and $\Sigma^+$ are ground states with charge, it will be easier to detect them in experiments.
\begin{table}
\caption{Masses and decay widths of the two predicted $N^*$ with hidden beauty ~\cite{plb709}.}
\label{decay-width}
\begin{tabular}{|c|cc|cc|}
\hline
 &$PB\rightarrow PB$ & &$PV\rightarrow PV$ &\\
\hline
 &$N^*(11052)$ & &$N^*(11100)$ &\\
\hline
M (MeV) &11052 & &11100 &\\
\hline
$\Gamma$ (MeV) &1.38 & &1.33 &\\
\hline
 &$\pi N$\ \ \  &0.10 &$\rho\pi$\ \ \  &0.09\\
 &$\eta N$\ \ \ &0.21 &$\omega N$\ \ \ &0.30\\
$\Gamma_i$ (MeV)&$\eta' N$\ \ \ &0.11 &$K^*\Sigma$\ \ \ &0.39\\
 &$K\Sigma$\ \ \ &0.42 &$\Upsilon N$\ \ \ &0.51\\
 &$\eta_b N$\ \ \ &0.52 & & \\
\hline
\end{tabular}
\end{table}

Within an effective Lagrangian approach and isobar model, the production of the $N^*(11052)$ resonance is investigated in this work. The results will not only provide valuable information for searching the neutral $N^*(11052)$ resonance, but also enable us to have a more comprehensive understanding of the properties of the super-heavy resonance. The COMPASS (Common Muon and Proton Apparatus for Structure and Spectroscopy) experiment~\cite{nimpr577} at CERN's Super Proton Synchrotron (SPS) with pion beam of $\simeq$ 280 GeV will be a good platform for searching the super-heavy resonance with hidden beauty, which may substantiates our numerical results in the future.

This paper is organized as following. After the introduction, the main ingredients and formalism in the neutral $N^*(11052)$ production processes are discussed in Sec. \uppercase\expandafter{\romannumeral2}. In Section \uppercase\expandafter{\romannumeral3}, numerical results are presented and discussed. Finally, a brief summary is given in Sec. \uppercase\expandafter{\romannumeral4}.

\begin{figure}
\centering
\subfigure[\ signal]{\includegraphics[width=0.23\textwidth]{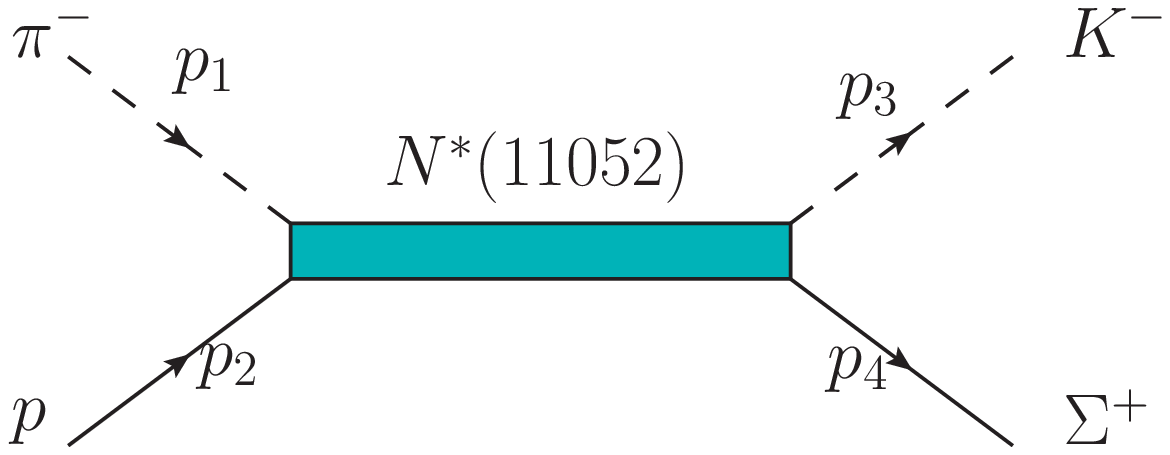}}
\subfigure[\ background]{\includegraphics[width=0.23\textwidth]{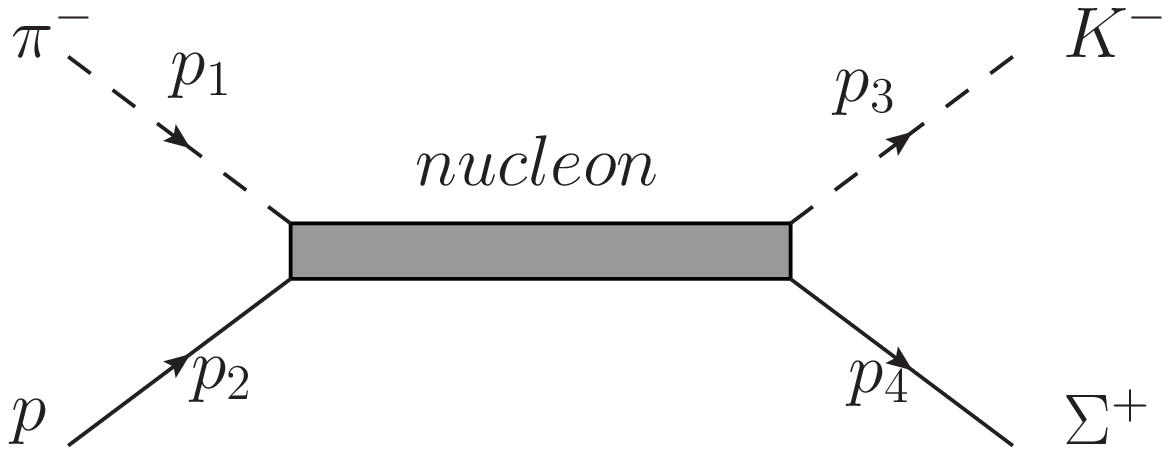}}
\caption{Feynman diagrams for the production the of $N^*(11052)$ resonance through $\pi^-p\rightarrow K^-\Sigma^+$ reaction, including the background of nucleon pole. Kinematical variables ($p_1,\ p_2,\ p_3,$ and $p_4$) used in our calculation are indicated in the figure.}
\label{feynman-diagram}
\end{figure}

\section{PRODUCTION OF neutral $N^*(11052)$ VIA $\pi^- p$ SCATTERING}

In the energy region of resonances, it is difficult to investigate the strong interaction at the quark-gluon level. Thus an effective Lagrangian approach in terms of hadrons is adopted, which is an important theoretical approach in investigating various processes~\cite{prc77,jpg41,epja51,epl109,prd92,prd93,ahep15,prdxy15,prd032,prc16,prche}. In this section, we introduce the theoretical formalism and ingredients to calculate the production process of the predicted $N^*(11052)$ resonance from two possible reactions, namely the $\pi^-p\rightarrow K^-\Sigma^+$ reaction and the $\pi^-p\rightarrow \eta_bn$ reaction.

\subsection{PRODUCTION FROM THE $\pi^-p\rightarrow K^-\Sigma^+$ reaction}

In Fig.~\ref{feynman-diagram}, the basic tree-level Feynman diagrams for the production of $N^*(11052)$ in the $\pi^-p\rightarrow K^-\Sigma^+$ reaction were shown, including background of nucleon pole. S-channel of $N^*(11052)$ as an intermediate state is considered.

The $N^*(11052)$ resonance, which is dynamically generated from the $PB$ channels, favors its spin-parity quantum number $J^p=(1/2)^-$~\cite{plb709}. The effective Lagrangian densities of the relevant interaction vertices can be written as following~\cite{plb709,prc67}:

\begin{align}
{\cal{L}}_{\pi NN^*}&=g_{\pi NN^*}\bar N\vec{\tau}\cdot\vec{\pi}N^*+H.c.,\\
{\cal{L}}_{K\Sigma N^*}&=g_{K\Sigma N^*}\bar N^*K\vec{\tau}\cdot\vec\Sigma+H.c.,\\
{\cal{L}}_{\pi NN}&=g_{\pi NN}\bar N\gamma_5\vec{\tau}\cdot\vec{\pi}N+H.c.,\\
{\cal{L}}_{K\Sigma N}&=g_{K\Sigma N}\bar N\gamma_5K\vec{\tau}\cdot\vec\Sigma+H.c..
\end{align}
where $N^*$ stands for the $N^*(11052)$ resonance and $\vec\tau$ stands for the Pauli matrix. The coupling constants $g_{\pi NN^*}$ and $g_{K\Sigma N^*}$ in the Lagrangian densities above are determined by the partial decay widths listed in Table~\ref{decay-width},

\begin{align}
\Gamma[N^*\rightarrow\pi N]&=\frac{3g^2_{\pi NN^*}}{4\pi}\frac{E_N+m_N}{m_{N^*}}|\vec{p_1}^{c.m.}|,\\
\Gamma[N^*\rightarrow K\Sigma]&=\frac{3g^2_{K\Sigma N^*}}{4\pi}\frac{E_{\Sigma}+m_\Sigma}{m_{N^*}}|\vec{p_2}^{c.m.}|,
\end{align}
with
\begin{align}
E_N&=\sqrt{|\vec{p_1}^{c.m.}|^2+m_N^2},\\
|\vec{p_1}^{c.m.}|&=\frac{\lambda^{\frac{1}{2}}(m^2_{N^*},\ m^2_{\pi},\ m^2_N)}{2m_{N^*}},\\
E_\Sigma&=\sqrt{|\vec{p_2}^{c.m.}|^2+m_\Sigma^2},\\
|\vec{p_2}^{c.m.}|&=\frac{\lambda^{\frac{1}{2}}(m^2_{N^*},\ m^2_K,\ m^2_\Sigma)}{2m_{N^*}},
\end{align}
where $\lambda$ is the K\"{a}llen function with $\lambda(x,\ y,\ z)=(x-y-z)^2-4yz$. The $\vec{p_1}^{c.m.}$ and $\vec{p_2}^{c.m.}$ are the three momentum of the either outgoing particle in the center of mass (c.m.) frame, whereas $E_N$ and $E_\Sigma$ are the energy of $N$ and $\Sigma$ in c.m.\ frame.

With the equations above, we obtain the coupling constants $g_{\pi NN^*}=0.011$ and $g_{K\Sigma N^*}=0.024$. For coupling constants of the background channel $g_{\pi NN}=13.45$ (obtained from $g_{\pi NN}^2/4\pi=14.4$) as used in Ref.~\cite{prc61,epja49,ctp63}, while $g_{K\Sigma N}$ is obtained by $SU(3)$ symmetry $g_{K\Sigma N}=g_{\pi NN}(1-2\alpha_1)=2.69$ with $\alpha_1=0.4$.

In addition, due to the fact that hadrons are not point-like particles, form factors are introduced as used in Ref.~\cite{prc58,prc59,prc72},
\begin{align}
{\cal F}_{N^*}&=\frac{\Lambda^4_{N^*}}{\Lambda^4_{N^*}+(q^2-M_{N^*}^2)^2},\label{form-factor}\\
{\cal F}_{N}&=\frac{\Lambda^4_{N}}{\Lambda^4_{N}+(q^2-M_{N}^2)^2},\label{from-factor-bg}
\end{align}
where $M_{N^*}$ ($M_N$) and $\Lambda_{N^*}$ ($\Lambda_N$) correspond to the mass and cut-off parameter of the intermediate $N^*(11052)$ resonance (nucleon pole).

Finally, for propagator $G_{N^*}(q)$, we adopt the Breit-Wigner formula,
\begin{align}\label{propagator}
G_{N^*}(q)=i\frac{q\!\!\! /+M_{N^*}}{q^2-M^2_{N^*}+iM_{N^*}\Gamma_{N^*}}
\end{align}
where $\Gamma_N^*$ stands for the full decay width of $N^*(11052)$. For the background of nucleon pole, its form can be written as:
\begin{align}\label{propagator-bg}
G_N(q)=i\frac{q\!\!\! /+M_N}{q^2-M^2_N}
\end{align}

With the ingredients above, the invariant amplitudes for the $\pi^-p\rightarrow K^-\Sigma^+$ reaction, according to the contributions shown in Fig~\ref{feynman-diagram}, can be written as:

\begin{align}
{\cal M}_{tot}=&{\cal M}_{N^*}+{\cal M}_{N},\\
{\cal M}_{N^*}=&2g_{\pi NN^*}g_{K\Sigma N^*}\bar u(p_4,s_4)G_{N^*}(q_s)u(p_2,s_2){\cal F}_{N^*}(q_s^2),\\
{\cal M}_{N}=&2g_{\pi NN}g_{K\Sigma N}\bar u(p_4)\gamma_5G_N(q_s)\gamma_5u(p_2){\cal F}_N(q_s^2),
\end{align}
where $q_s=p_1+p_2$, $s_2$ and $s_4$ are the polarization variables of the initial proton and the final $\Lambda$ baryon.

The unpolarized differential cross section in the center of mass frame for the $\pi^-p\rightarrow K^-\Sigma^+$ reaction can be derived from the invariant amplitude square $|{\cal M}|^2$, read as:
\begin{align}\label{dcs}
\frac{d\sigma}{dcos\theta}=\frac{4m_pm_\Lambda}{32\pi^2 s}\frac{|\vec{p_3}^{c.m.}|}{|\vec{p_1}^{c.m.}|}\bigg(\frac{1}{2}\sum_{s_2,s_4}|{\cal M}|^2\bigg),
\end{align}
where $s=W^2$ is total energy squared in c.m.\ frame, $\theta$ denotes the angle of the outgoing $K^-$ meson relative to the $\pi^-$ beam direction in c.m.\ frame.

\subsection{PRODUCTION FROM THE $\pi^-p\rightarrow \eta_bn$ REACTION}

\begin{figure}
\centering
\subfigure[\ s-channel signal]{\includegraphics[width=0.23\textwidth]{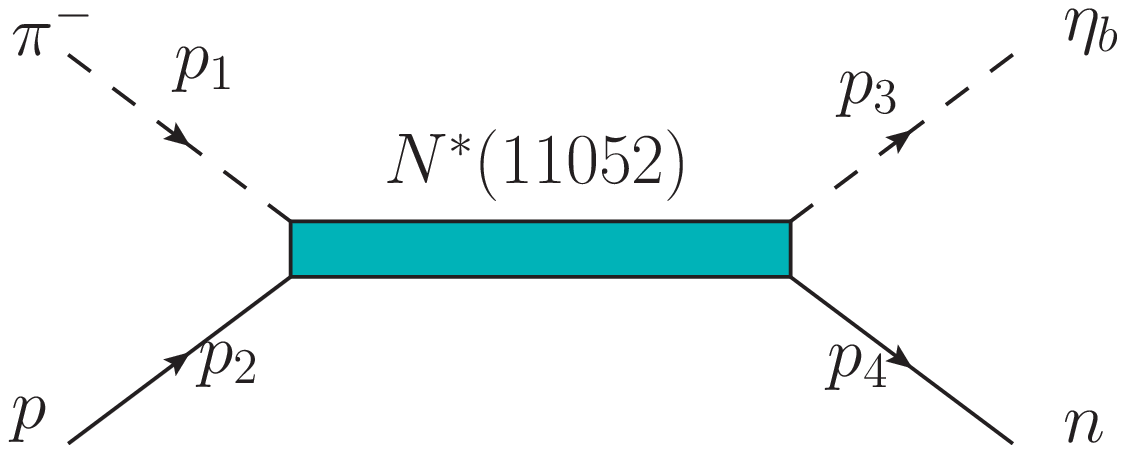}}
\subfigure[\ u-channel signal]{\includegraphics[width=0.23\textwidth]{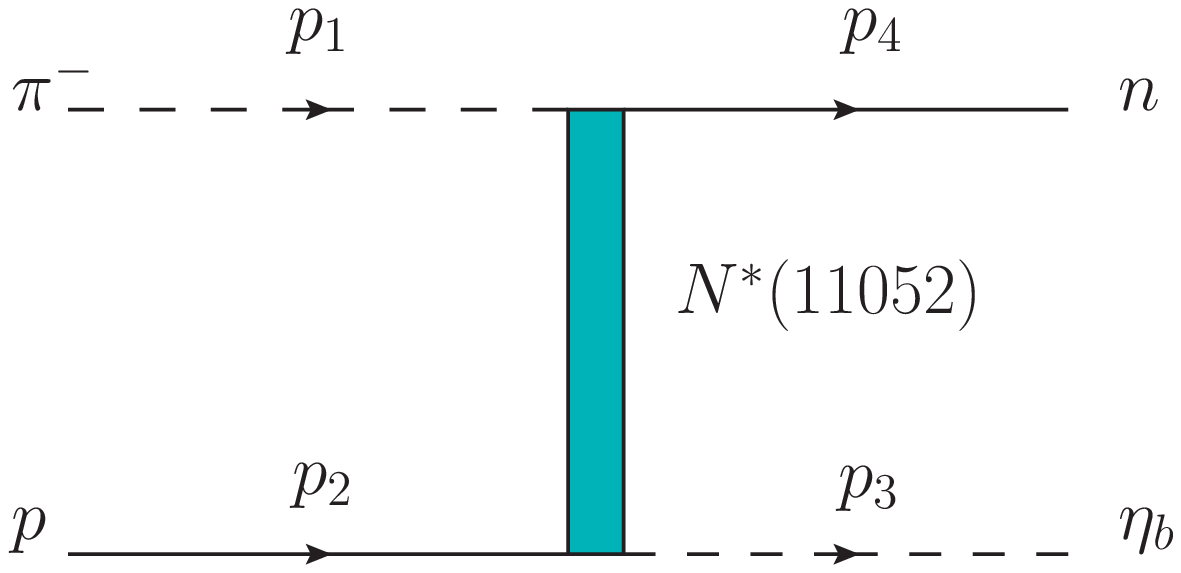}}
\subfigure[\ s-channel background]{\includegraphics[width=0.23\textwidth]{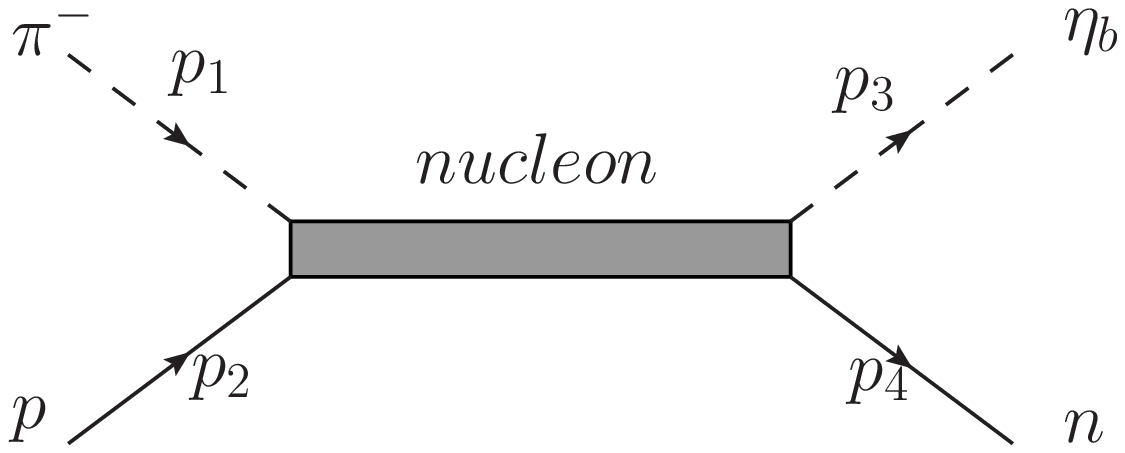}}
\subfigure[\ u-channel background]{\includegraphics[width=0.23\textwidth]{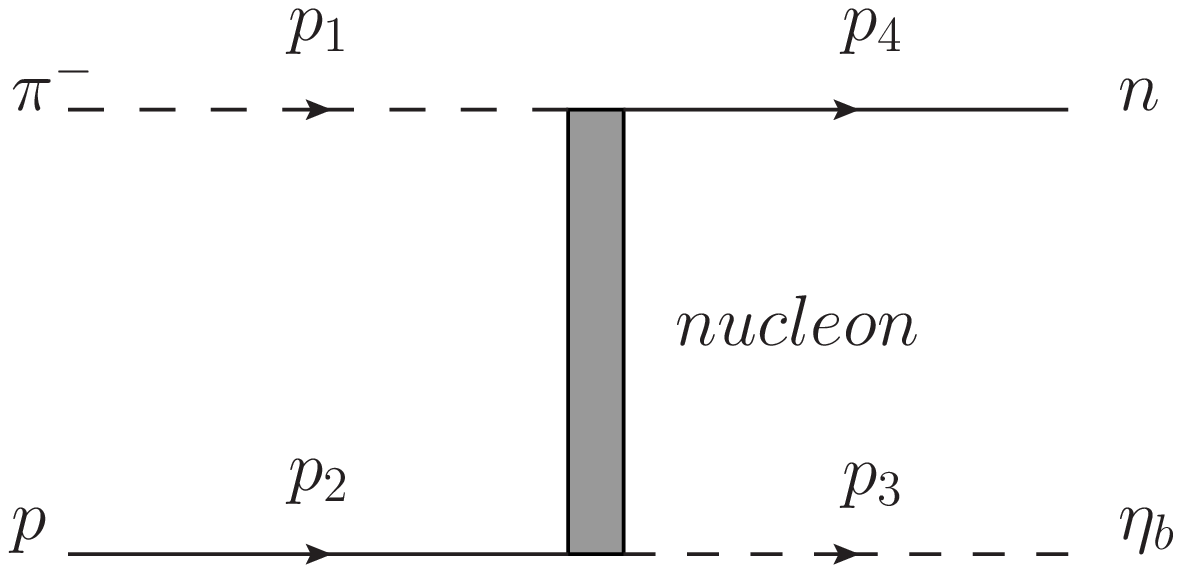}}
\caption{Feynman diagrams for the $\pi^-p\rightarrow \eta_bn$ reaction including s-channel and u-channel, with background of nucleon pole considered. Kinematical variables ($p_1,\ p_2,\ p_3,$ and $p_4$) used in our calculation are indicated in the figure.}
\label{feynman-diagram-etabN}
\end{figure}

In table~\ref{decay-width} there is another channel suitable for searching the $N^*(11052)$ resonance, the $\eta_b N$ decay channel because of its relative big decay branch ratio. The tree-level Feynman diagrams of the $\pi^-p\rightarrow \eta_bn$ reaction are shown in Fig.~\ref{feynman-diagram-etabN}, with s-channel and u-channel considered.

The Lagrangians for the relevant vertices in Fig.~\ref{feynman-diagram-etabN} can be written as~\cite{plb709,prc67},

\begin{align}
{\cal{L}}_{\pi NN^*}&=g_{\pi NN^*}\bar N\vec{\tau}\cdot\vec{\pi}N^*,\\
{\cal{L}}_{\eta_bNN^*}&=g_{\eta_bNN^*}\bar N^*\eta_bN+H.c.,\\
{\cal{L}}_{\pi NN}&=g_{\pi NN}\bar N\gamma_5\vec{\tau}\cdot\vec{\pi}N,\\
{\cal{L}}_{\eta_bNN}&=g_{\eta_bNN}\bar N\gamma_5\eta_bN+H.c..
\end{align}

The coupling constants are taken as $g_{N^*N\pi}^2/4\pi=1.03\times 10^{-5}$, $g_{N^*N\eta_b}^2/4\pi=1.81\times 10^{-3}$, $g_{\pi NN}=14.4$ and $g_{pp\eta_b}=1\times 10^{-6}$ as used in Ref.~\cite{plb709}.

For form factors and propagators, we adopt the same form of Eqs.~\eqref{form-factor}~\eqref{from-factor-bg}~\eqref{propagator}~\eqref{propagator-bg}. The total invariant amplitudes for the $\pi^-p\rightarrow \eta_bn$ reaction can be written as:

\begin{align}
{\cal M}_{tot}=&{\cal M}_{N^*}+{\cal M}_N,\\
{\cal M}_{N^*}=&{\cal M}_{N^*}^s+{\cal M}_{N^*}^u,\\
{\cal M}_{N^*}^s=&\sqrt2g_{\pi NN^*}g_{\eta_bNN^*}\bar u(p_4,s_4)G_N^*(q_s)\\\notag
&\times u(p_2,s_2){\cal F}_{N^*}(q_s^2),\\
{\cal M}_{N^*}^u=&\sqrt2g_{\pi NN^*}g_{\eta_bNN^*}\bar u(p_4,s_4)G_N^*(q_u)\\\notag
&\times u(p_2,s_2){\cal F}_{N^*}(q_u^2),\\
{\cal M}_N=&{\cal M}_N^s+{\cal M}_N^u,\\
{\cal M}_N^s=&\sqrt2g_{\pi NN}g_{\eta_bNN}\bar u(p_4,s_4)\gamma_5G_N(q_s)\\\notag
&\times\gamma_5u(p_2,s_2){\cal F}_N(q_s^2),\\
{\cal M}_N^u=&\sqrt2g_{\pi NN}g_{\eta_bNN}\bar u(p_4,s_4)\gamma_5G_N(q_u)\\\notag
&\times\gamma_5u(p_2,s_2){\cal F}_N(q_u^2),
\end{align}
where $q_s=p_1+p_2$ and $q_u=p_1-p_3$. $s_2$ and $s_4$ are the polarization variables of the initial proton and final neutron.

The unpolarized differential cross section derived from the invariant amplitude square $|{\cal M}|^2$ can be written as:
\begin{align}\label{dcs}
\frac{d\sigma}{dcos\theta}=\frac{4m_pm_n}{32\pi^2 s}\frac{|\vec{p_3}^{c.m.}|}{|\vec{p_1}^{c.m.}|}\bigg(\frac{1}{2}\sum_{s_2,s_4}|{\cal M}|^2\bigg).
\end{align}

\section{NUMERICAL RESULTS AND DISCUSSION}

With the formalism and ingredients given above, cross sections for the neutral $N^*(11052)$ production processes can be evaluated. The results for two production processes are discussed in the following two subsections.

\subsection{$\pi^-p\rightarrow K^-\Sigma^+$}
In this subsection, we discuss the numerical results for the $\pi^-p\rightarrow K^-\Sigma^+$ production process, with total and differential cross sections given.

\begin{figure}
\centering
\subfigure[]{\includegraphics[width=0.45\textwidth]{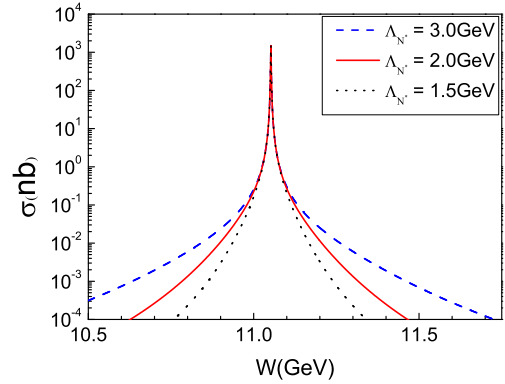}}
\subfigure[]{\includegraphics[width=0.45\textwidth]{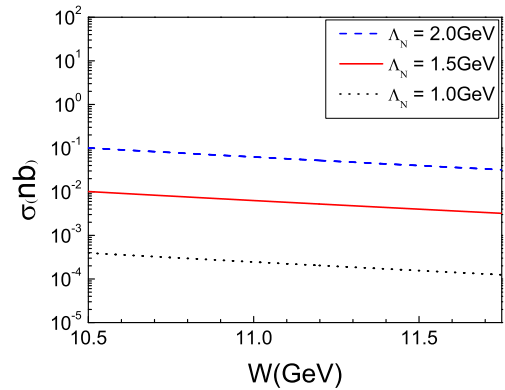}}
\subfigure[]{\includegraphics[width=0.45\textwidth]{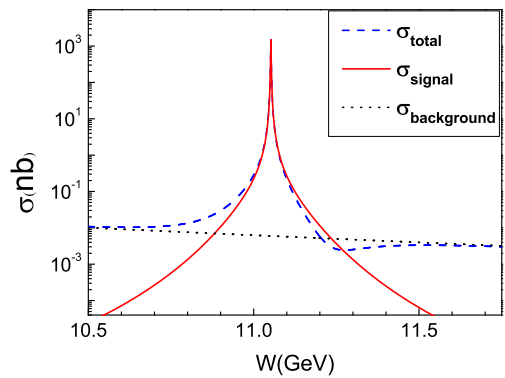}}
\caption{(Color online) Total cross section versus c.m.\ energy $W$ for the $\pi^-p\rightarrow K^-\Sigma^+$ reaction. (a) Signal with different cut-off parameters $\Lambda_{N^*}=$ 1.5, 2.0 and 3.0 GeV. (b) Background with different cut-off parameters $\Lambda_N=$ 1.0, 1.5 and 2.0 GeV. (c) Contributions from the signal and the background, with cut-off parameters set at $\Lambda_{N^*}=2.0$ GeV and $\Lambda_{N}=1.5$ GeV. The ``total" denotes the contributions from both the signal and the background.}
\label{KSigma-tcs}
\end{figure}
In Fig.~\ref{KSigma-tcs} (a), the signal cross section with variations of different cut-off parameters $\Lambda_{N^*}=$ 1.5, 2.0 and 3.0 GeV are shown. There is a significant enhancement at c.m.\ energy $W=11.05$ GeV. The cut-off parameters determine the widths of the peaks, but the positions of the peaks are independent of the cut-off parameters. These properties originate from form factors Eq.~\eqref{form-factor}, form factors reach maximize value of 1 when $q^2=M_{N^*}^2$.

In Fig.~\ref{KSigma-tcs} (b), the background is shown, with different cut-off parameters $\Lambda_N=$ 1.0, 1.5 and 2.0 GeV. Although, the values of the cut-off parameter $\Lambda_N$ have a considerable effect on the magnitude of the background cross section, it is still negligible when compared with the signal around c.m.\ energy $W=11$ GeV. Because this energy region $W\simeq 11$ GeV is too far away from the mass of nucleon, form factor of the nucleon pole Eq.~\eqref{from-factor-bg} is rather small, so the background is suppressed.

In Fig.~\ref{KSigma-tcs} (c), with cut-off parameters set at $\Lambda_{N^*}=2.0$  GeV and $\Lambda_N=1.0$ GeV, the comparison of the signal with the background, together with the total (including signal and background) channel are shown. The signal channel is absolutely dominant near c.m.\ energy $W=11.05$ GeV, we conclude that 11 GeV $\simeq W\simeq$ 11.1 GeV will be an ideal energy region for searching the $N^*(11052)$ resonance through the $\pi^-p\rightarrow K^-\Sigma^+$ reaction. In the laboratory frame, the incident pion beam momentum corresponds to such c.m.\ energy $W=11$ GeV is $p_{lab}=64$ GeV, which can be done by the COMPASS experiment~\cite{nimpr577}. It is worth to mention that beyond $W=11.1$ GeV, the interference of the signal and the background has a considerable cancelling effect, the total channel is smaller than the signal or the background.

\begin{figure}
\centering
\includegraphics[width=0.45\textwidth]{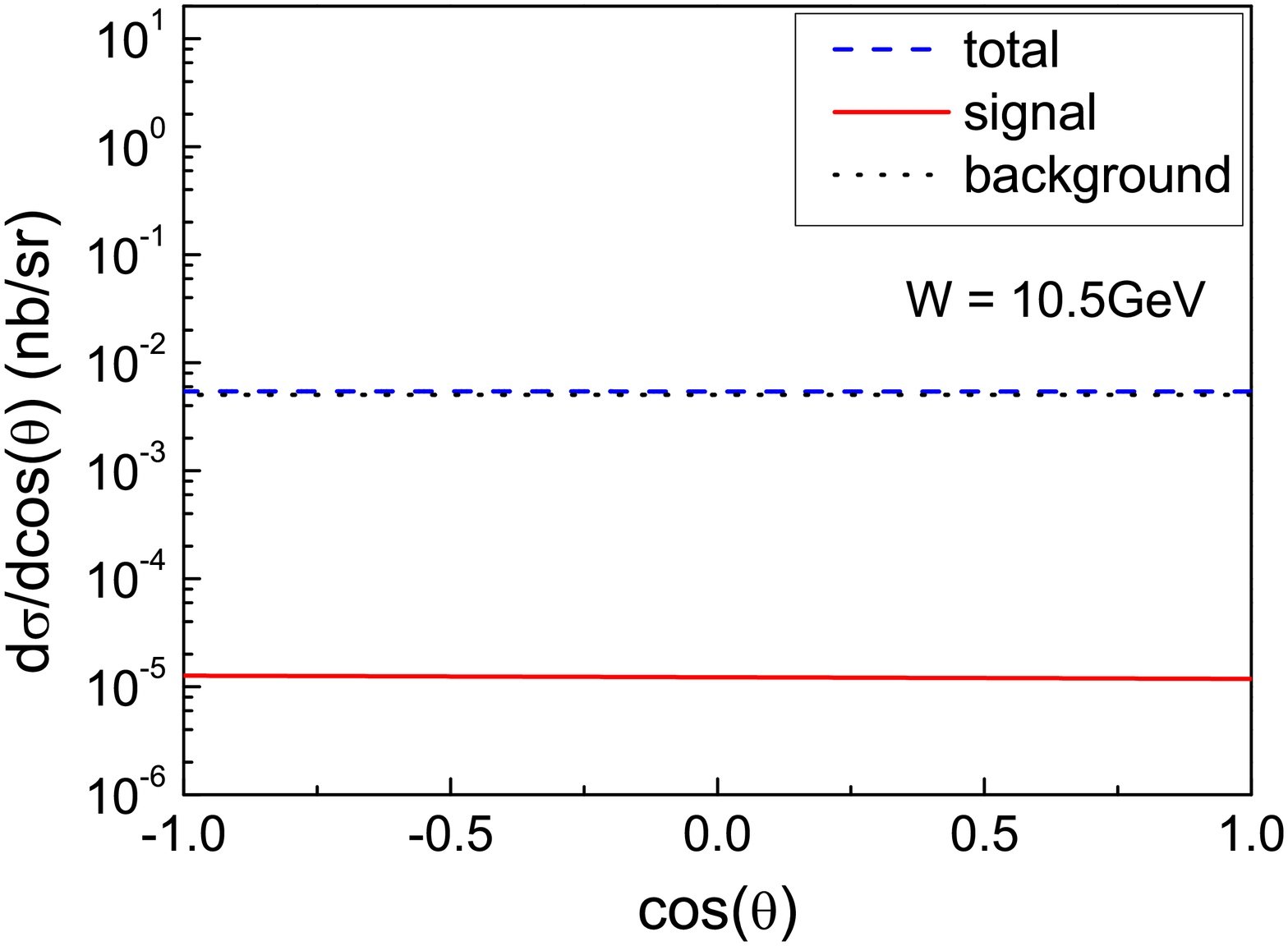}
\includegraphics[width=0.45\textwidth]{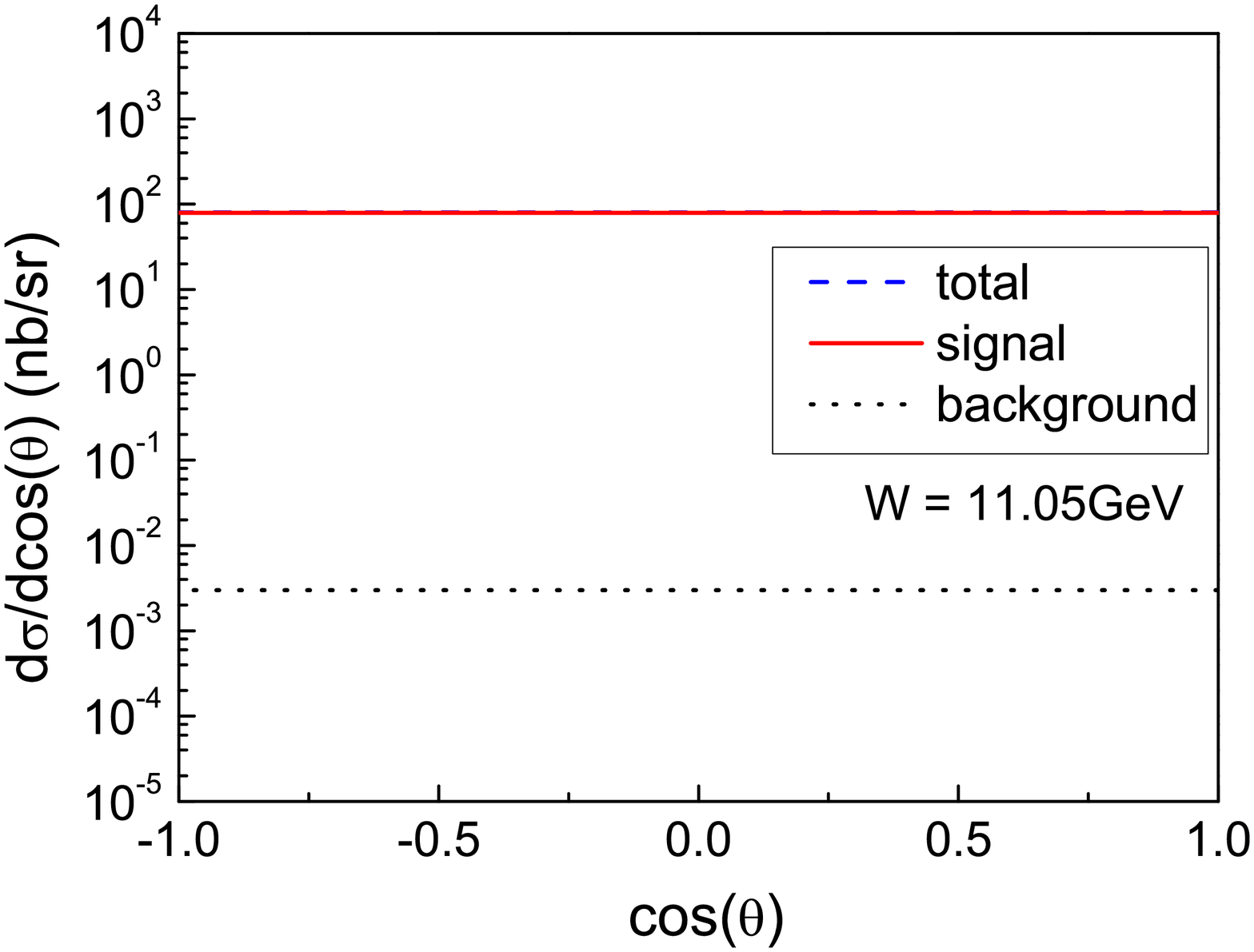} 
\includegraphics[width=0.45\textwidth]{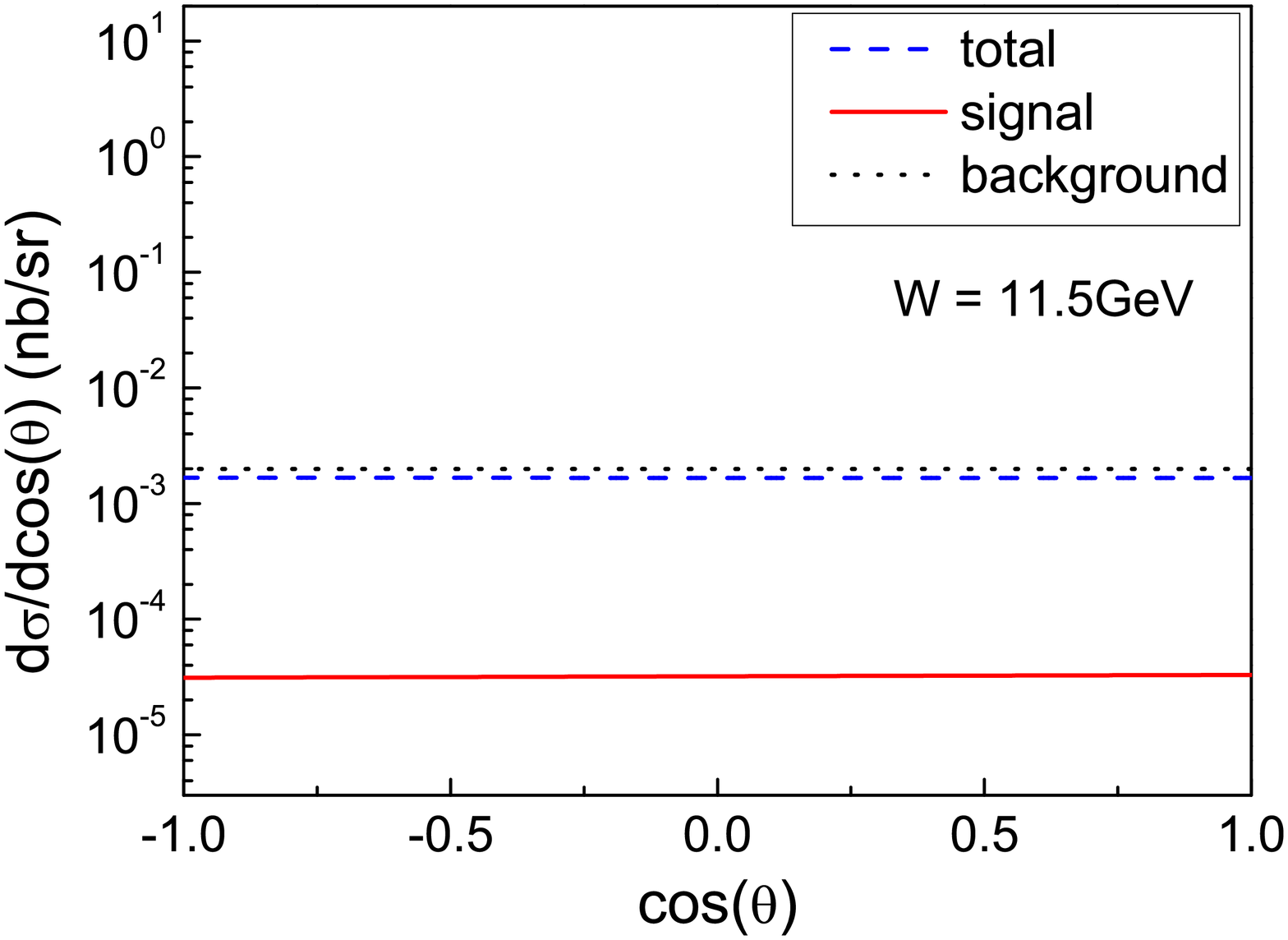}
\caption{(Color online) The differential cross section for the $\pi^-p\rightarrow K^-\Sigma^+$ reaction, at c.m.\ energies $W=$ 10.5, 11.05 and 11.5 GeV. Here the ``total" denotes the differential cross section including contributions from both the signal and the background. The red solid line stands for the signal, the black dotted line stands for the background and the blue dashed line stands for the total contribution.}
\label{KSigma-dcs}
\end{figure}
Fig.~\ref{KSigma-dcs} shows the differential cross sections of the signal, background and total channel, at center of mass energy $W=10.5$, 11.05 and 11.5 GeV. The contribution from signal are dominant at c.m.\ energies $W=11.05$ GeV, whereas when $W$ deviate from the threshold of $N^*(11052)$, e.g., $W=10.5$ GeV or $W=11.5$ GeV the contribution from the background becomes significant. Differential cross sections are independent of the outgoing angle $\theta$, for these are s-channel reactions. All these properties can be tested by future experiment.

\subsection{$\pi^- p\rightarrow \eta_b N$}
In this subsection, we discuss the numerical results for the $\pi^- p\rightarrow \eta_b N$ production process, with total and differential cross sections given.

\begin{figure}
\centering
\subfigure[]{\includegraphics[width=0.45\textwidth]{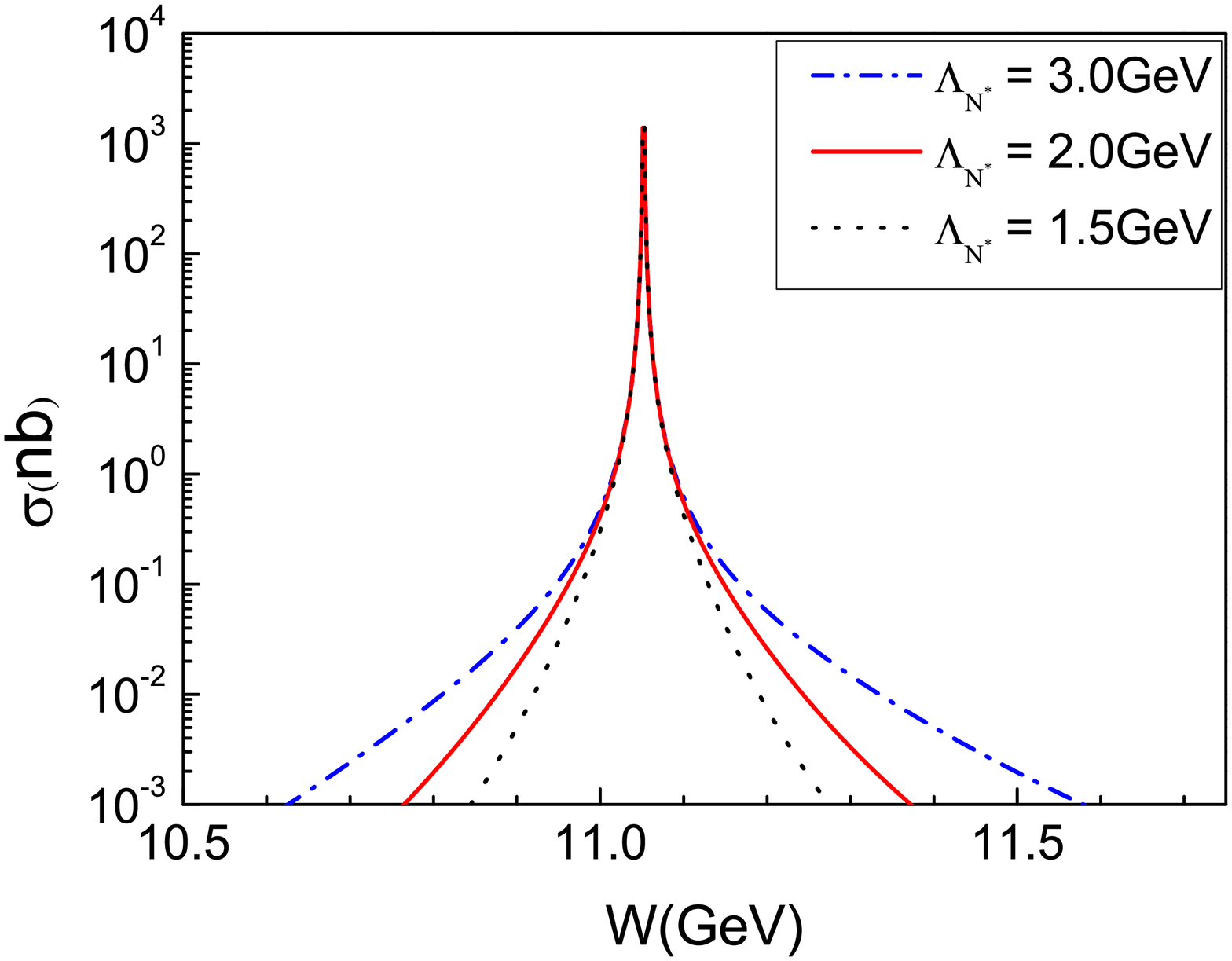}}
\subfigure[]{\includegraphics[width=0.45\textwidth]{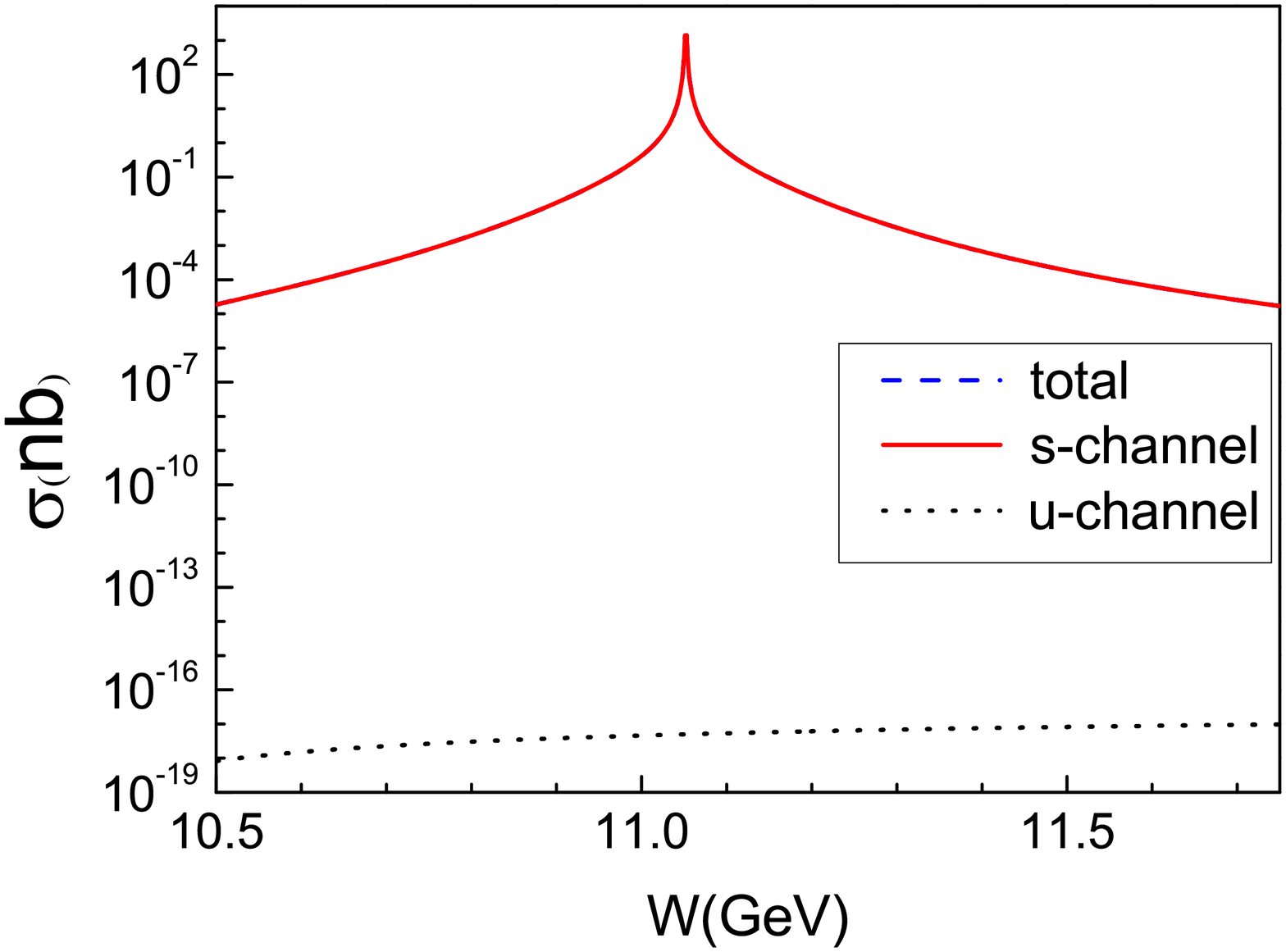}}
\subfigure[]{\includegraphics[width=0.45\textwidth]{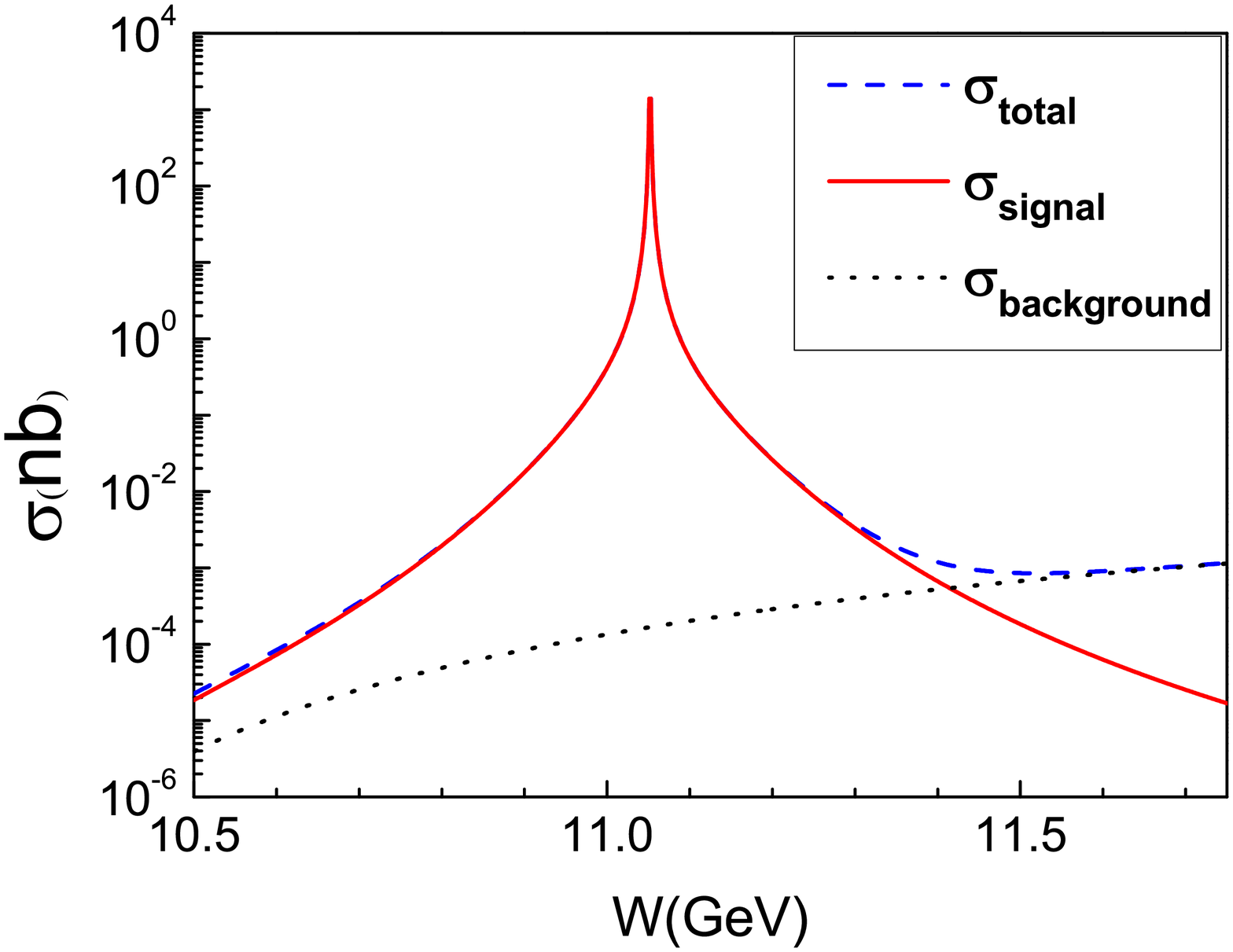}}
\caption{(Color online)(a) Total cross section of the signal channel, with cut-off parameters $\Lambda_{N^*}=$ 1.5, 2.0 and 3.0 GeV. (b) Comparison of s-channel with u-channel, with cut-off parameter sets at $\Lambda_{N^*}=2.0$ GeV. The ``total" denotes the contributions from both s-channel and u-channel. (c) Comparison of the signal with the background, with cut-off parameters set at $\Lambda_{N^*}=2.0$ GeV and $\Lambda_{N}=1.5$ GeV. The ``total" denotes the contributions from both signal and background.}
\label{etabN-tcs-signal}
\end{figure}
Fig.~\ref{etabN-tcs-signal} (a) shows the total cross section of the signal channel, with variations of different cut-off parameters $\Lambda_{N^*}=$ 1.5, 2.0 and 3.0 GeV. The comparison of the contributions from s-channel and u-channel are shown in Fig.~\ref{etabN-tcs-signal} (b). Although the u-channel reaction is included in this production process, its contribution is rare as shown in Fig.~\ref{etabN-tcs-signal} (b). Because the $N^*(11052)$ resonance as a exchange particle is too heavy, the u-channel reaction is deeply suppressed. In Fig.~\ref{etabN-tcs-signal} (c) the comparison of the signal and the background is shown, contribution from the background becomes significant when the c.m.\ energy goes beyond 11.5 GeV. But in the energy region of 11 GeV $\lesssim W \lesssim$ 11.1 GeV, the signal is absolutely dominant, so we conclude this is a best energy window for searching the $N^*(11052)$ resonance in our calculation. And the peaks can reach up to the magnitude of 1 $\mu b$ around c.m. energy $W\simeq 11.052$ GeV. It is about the same magnitude as in the neutral $p_c$ pentaquark states production process in $\pi^-p\rightarrow J/\psi n$ reaction~\cite{prd93}, which is appreciable to be observed and measured in experiment.

\begin{figure}[ht]
\centering
\includegraphics[width=0.45\textwidth]{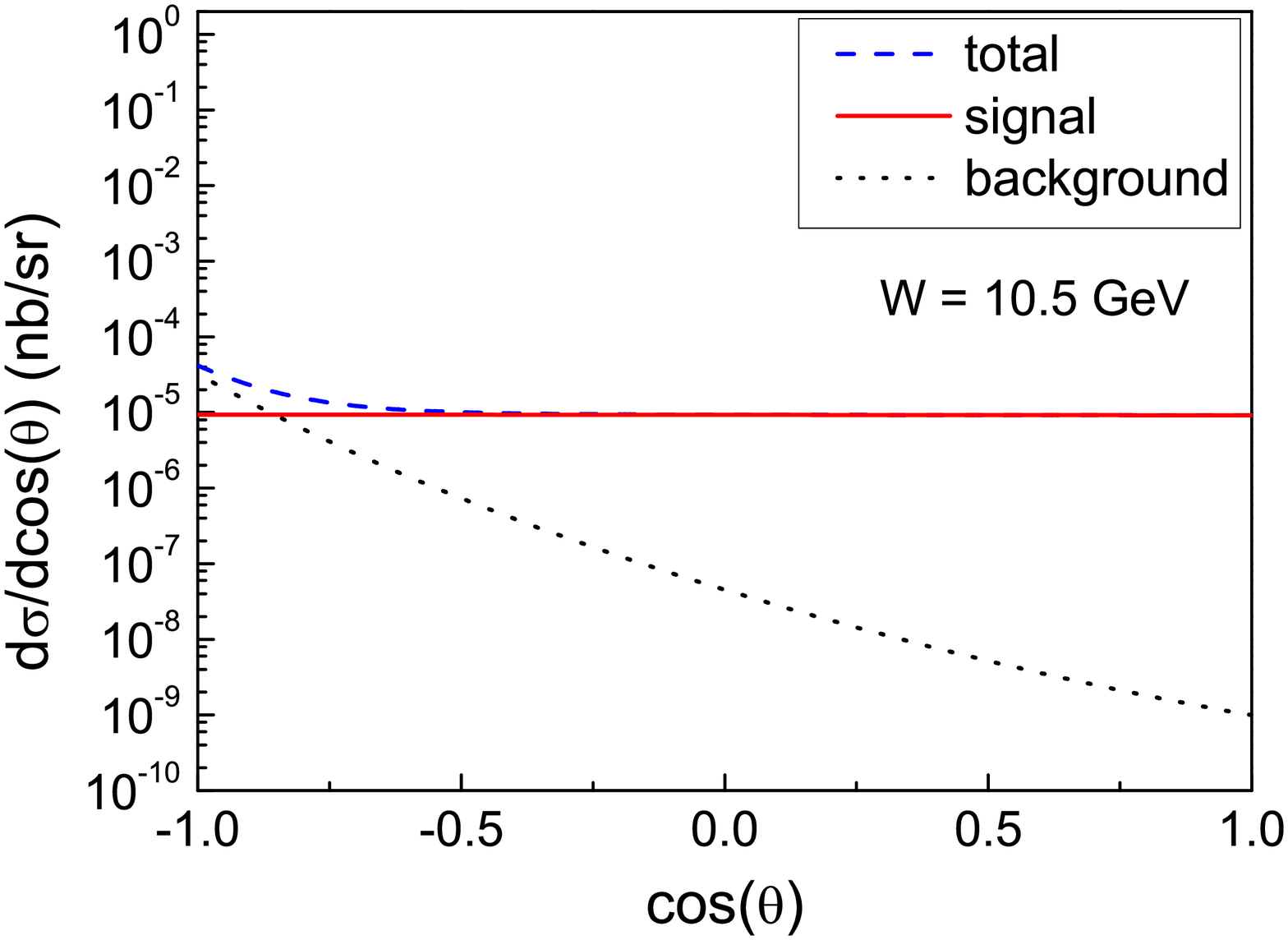}
\includegraphics[width=0.45\textwidth]{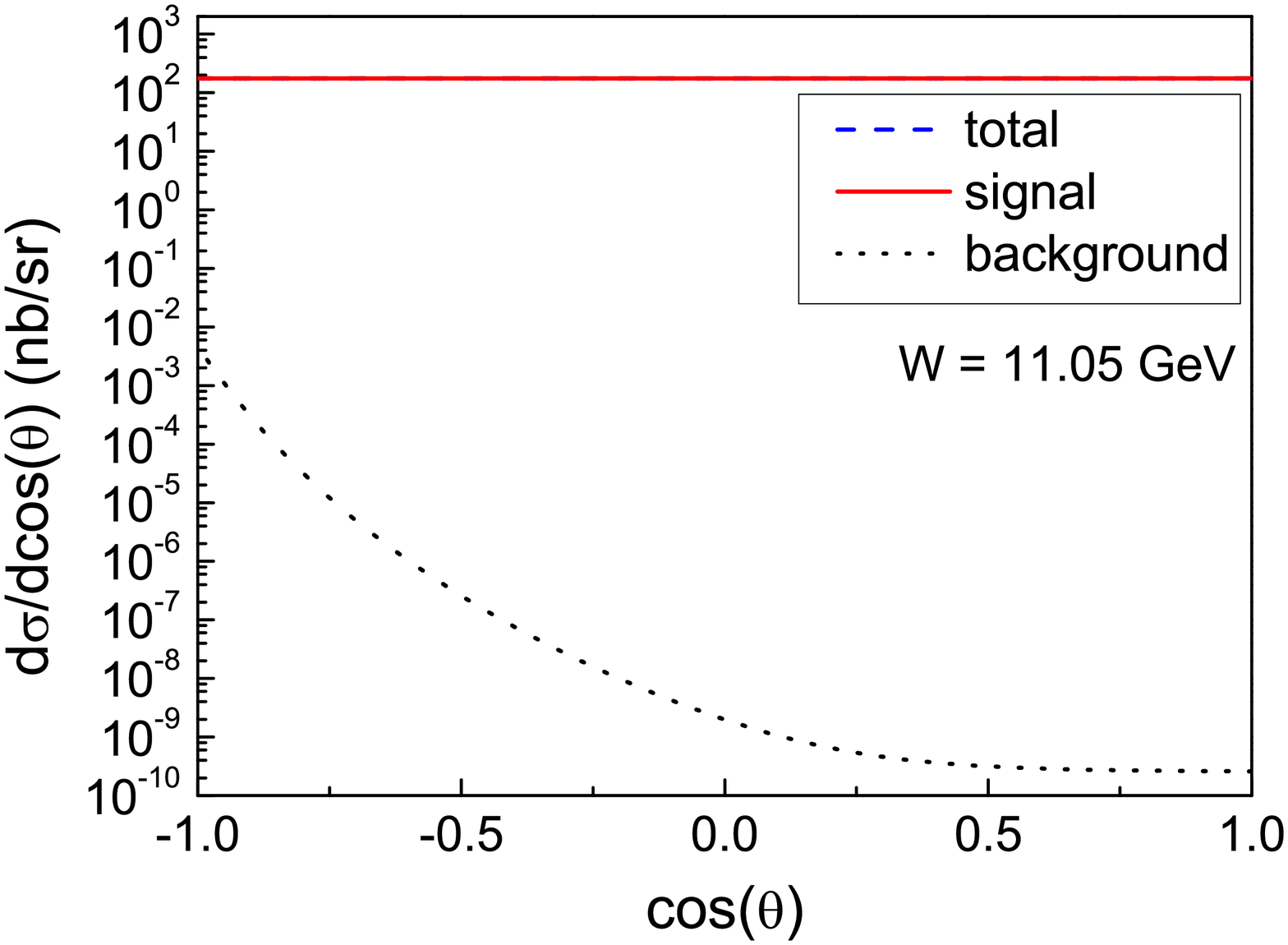}
\includegraphics[width=0.45\textwidth]{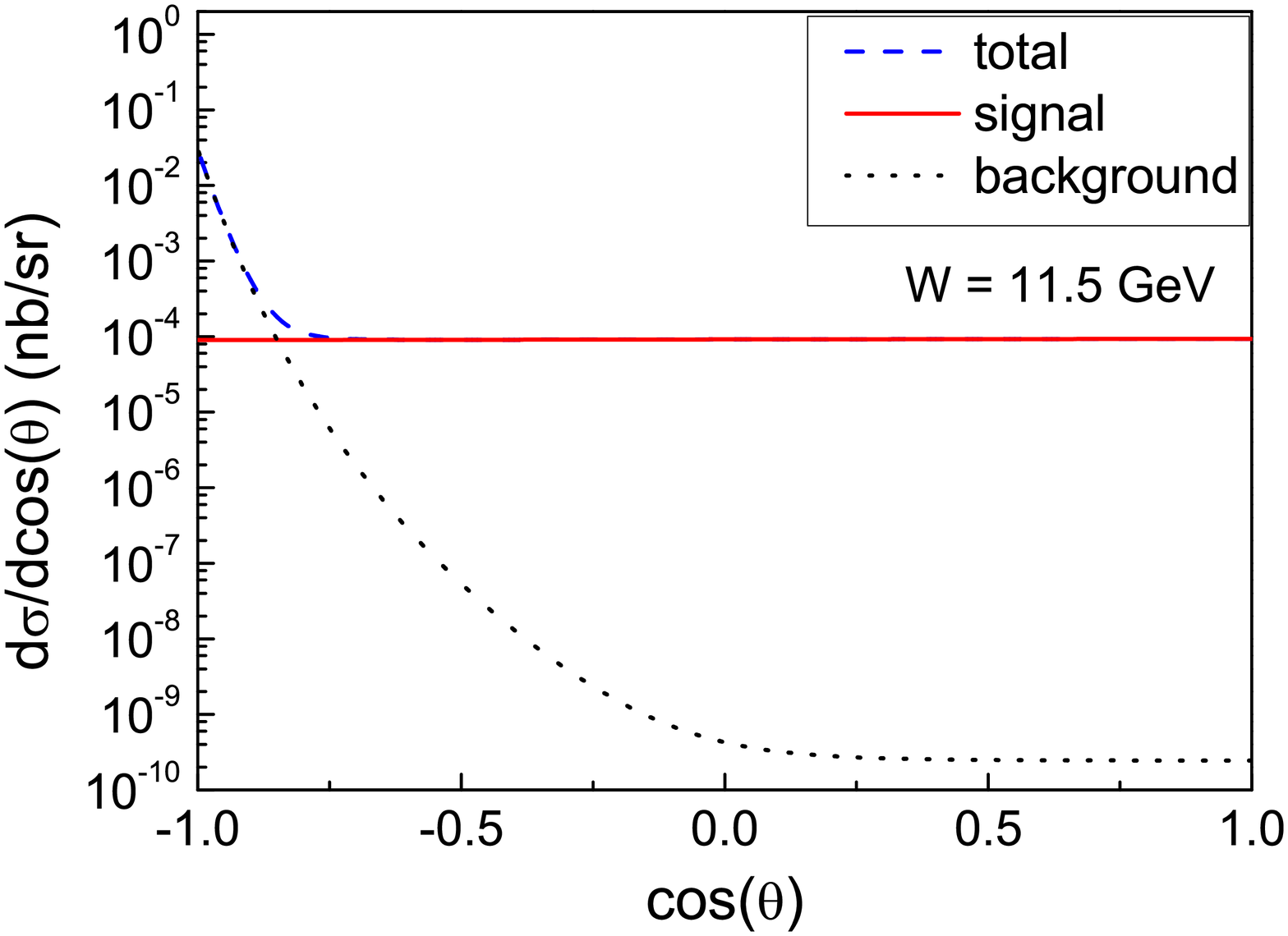}
\caption{(Color online) The differential cross section of the $\pi^- p\rightarrow \eta_b N$ reaction at c.m.\ energies $W=10.5$, 11.05 and 11.5 GeV. The ``total" denotes the differential cross section including contributions from both the signal and the background. The red solid line stands for the signal, the black dotted line stands for the background and the blue dashed line stands for the total contribution.}
\label{etabN-dcs}
\end{figure}
The differential cross section of the $\pi^- p\rightarrow \eta_b N$ reaction is shown in Fig.~\ref{etabN-dcs}, at center of energy energy of $W=10.5$, 11.05 and 11.5 GeV. For the signal channel, differential cross section is independent of the outgoing angle $\theta$ since the u-channel rarely contributes. But for the background channel there is a relative large contribution from the backward angle as shown in Fig.~\ref{etabN-dcs}, it originates from the u-channel contribution of nucleon pole exchange. These properties are expected to be tested by further experiments.
\section{SUMMARY}

In this work, we investigated the discovery potential of the neutral $N^*(11052)$ resonance through $\pi^- p$ scattering within an effective Lagrangian approach. Two possible reactions $\pi^-p\rightarrow K^-\Sigma^+$ and $\pi^-p\rightarrow \eta_bn$ are studied, with the background of nucleon pole considered. The results showed there is a significant enhancement for the production of $N^*(11052)$ resonance around c.m.\ energy $W=11.05$ GeV. Moreover, it is found that in the $\pi^-p\rightarrow \eta_bn$ reaction s-channel mediated by the $N^*(11052)$ is absolutely dominant, whereas the contribution from the u-channel $N^*(11052)$ exchange is negligible. The $\pi^-p\rightarrow K^-\Sigma^+$ reaction is especially ideal for searching the $N^*(11052)$ resonance, for the final products $K^-$ and $\Sigma^+$ are ground states with charge, it will be easier to detect them in experiment.

It is worth to mention that the $N^*(11052)$ resonance is a predicted state~\cite{plb709}, have not been observed by experiments or confirmed by other evidence. Its existence and many properties remain in doubt, nevertheless, this work constitutes a first step in this direction, more works concerning these exotic resonances with hidden beauty are needed. Although, the model we employed in this work depends strongly on the cut-off parameters within the form factors, the magnitudes and the positions of the signal peaks do not vary with different parameter values. The background analysis is less than exhaustive, due to the fact that the experimental data and theoretical study is scarce in this hidden beauty energy region. We investigated the contribution from the nucleon pole background as an illumination, it is shown that near the threshold the contribution from the nucleon pole is negligible when compared with the signal's. If this possible penta-quark candidate state really exist, according to our research in this work, it will be very likely to detect such resonance through the $\pi^- p$ scattering near the production threshold.

The center of mass energy $W\simeq$ 11-11.1 GeV would be a best energy window for searching the $N^*(11052)$ resonance in the $\pi^- p$ scattering discussed above. Total cross section can reach up to the magnitude of 1 $\mu b$ near the threshold of the $N^*(11052)$ resonance. The results will not only provides us a feasible way for searching the super-heavy resonance $N^*(11052)$, but also enable us to have a more comprehensive understanding of the properties of the super-heavy resonance. The COMPASS experiment will be a good platform for conducting such experiments. The estimated total cross sections, together with the angular distributions, can be checked by future COMPASS experiments.

\section{Conflict of Interests}
The authors declare that there is no conflict of interests
regarding the publication of this paper.
\section{Acknowledgments}
One of the authors (X.Y.W) gratefully acknowledges discussions with Alexey Guskov. C.C. acknowledges Ju-Jun Xie for his valuable help.


\begin{thebibliography}{90}

\bibitem{prd18} N. Isgur and G. Karl, Phys. Rev. D {\bf 18} 4187 (1978).

\bibitem{prd34} S. Capstick and N. Isgur, Phys. Rev. D {\bf 34} 2809 (1986).

\bibitem{csb59} X. Liu, Chin. Sci. Bull. {\bf 59} 3815 (2014).

\bibitem{epjc71} N. Brambilla et al., Eur. Phys. J. C {\bf 71} 1534 (2011).

\bibitem{ppnp45} S. Capstick and W. Roberts, Prog. Part. Nucl. Phys. {\bf 45} S241 (2000).

\bibitem{npa675} H.B. Li et al. (BES Collaboration), Nucl. Phys. A {\bf 675} 189c (2000); M. Ablikim et al. (BES Collaboration), hepex/0405030.

\bibitem{plb510} J.Z. Bai et al. (BES Collaboration), Phys. Lett. B {\bf 510} 75 (2001).

\bibitem{pn16} B.S. Zou et al. (BES Collaboration), PiN Newsletter {\bf 16} 174 (2002).

\bibitem{ijmpa20} H.X. Yang et al. (BES Collaboration), Int. J. Mod. Phys. A {\bf 20} 1985 (2005).

\bibitem{prl96} B.C. Liu and B. S. Zou, Phys. Rev. Lett. {\bf 96} 042002 (2006).

\bibitem{prc77} J.J. Xie, C. Wilkin, B.S. Zou, Phys. Rev. C {\bf 77} 058202 (2008).

\bibitem{npa669} Helminen C. and Riska D. O., Nucl. Phys. A {\bf 699} 624 (2002).

\bibitem{npa835} Zou B. S., Nucl. Phys. A {\bf 835} 199 (2010).

\bibitem{prl105} J.J. Wu, R. Molina, E. Oset, B. S. Zou, Phys. Rev. Lett. {\bf 105} 232001 (2010).

\bibitem{prc84} J.J. Wu, R. Molina, E. Oset, B. S. Zou, Phys. Rev. C {\bf 84} 015203 (2011).

\bibitem{jpg41} Y. Huang, J. He, H. F. Zhang and X. R. Chen, J. Phys. G {\bf 41} 115004 (2014).

\bibitem{epja51} X.Y. Wang and X. R. Chen, Eur. Phys. J. A {\bf 51} 85 (2015).

\bibitem{epl109} X.Y. Wang and X. R. Chen, Eur. Phys. Lett. {\bf 109} 41001 (2015).

\bibitem{prl115} R. Aaij et al. (LHCb Collaboration), Phys. Rev. Lett. {\bf 115} 072001 (2015).

\bibitem{prd92} Q. Wang, X. H. Liu and Q. Zhao, Phys. Rev. D {\bf 92} 034022 (2015).

\bibitem{prd93} Q.F. L¨¹, X.Y. Wang, J.J. Xie, X.R. Chen and Y.B. Dong, Phys. Rev. D {\bf 93} 034009 (2016).

\bibitem{prl108} A. Bondar et al. (Belle Collaboration), Phys. Rev. Lett. {\bf 108} 122001 (2012).

\bibitem{plb709} J.J. Wu, Lu Zhao and B. S. Zou, Phys. Lett. B {\bf 709} 70 (2012).

\bibitem{nimpr577} P. Abbon et al. (COMPASS Collaboration), Nucl. Instrum. Methods Phys. Res., Sect. A {\bf 577} 455 (2007).

\bibitem{ahep15} X.Y. Wang and X.R. Chen, Adv. High Energy Phys. 2015, 918231 (2015).

\bibitem{prdxy15} X.Y. Wang, X.R. Chen and A. Guskov, Phys. Rev. D {\bf 92} 094017 (2015).

\bibitem{prd032} X.Y. Wang and A. Guskov, Phys. Rev. D {\bf 92} 094032 (2015).

\bibitem{prc16} X. Y. Wang, J. He and H. Haberzettl, Phys. Rev. C {\bf 93} 045204 (2016).

\bibitem{prche} X. Y. Wang and J. He, Phys. Rev. C {\bf 93} 035202 (2016).

\bibitem{prc67} B.S. Zou, F. Hussain, Phys. Rev. C {\bf 67} 015204 (2003).

\bibitem{prc61} Lin Z. W., Ko C. M. and Zhang B., Phys. Rev. C {\bf 61} 024904 (2000).

\bibitem{epja49} D. Ronchen, M. Doring, F. Huang, H. Haberzettl, J. Haidenbauer, C. Hanhart, S. Krewald, U.G. Meissner, et al, Eur. Phys. J. A {\bf 49} 44 (2013).

\bibitem{ctp63} C.Z. Wu, Q.F. L\"{u}, J.J. Xie and X.R. Chen, Commun. Theo. Phys. {\bf 63} 215 {2015}

\bibitem{prc58} Feuster T. and Mosel U., Phys. Rev. C {\bf 58} 457 (1998).

\bibitem{prc59} Feuster T. and Mosel U., Phys. Rev. C {\bf 59} 460 (1999).

\bibitem{prc72} Shklyar V., Lenske H. and Mosel U., Phys. Rev. C {\bf 72} 015210 (2005).


\end{thebibliography}
\end{document}